\documentclass[aps,amsmath,showpacs,twocolumn]{revtex4}
\usepackage{graphicx}

\begin{document}

\title{The Real Message in the Sky}

\author{Douglas Scott} \email{dscott@phas.ubc.ca}
\author{J. P. Zibin} \email{zibin@phas.ubc.ca}
\affiliation{Department of Physics \& Astronomy\\
University of British Columbia,
Vancouver, BC, V6T 1Z1  Canada}

\begin{abstract}
A recent paper by Hsu \& Zee (physics/0510102) suggests that if a Creator
wanted to leave a message for us, and she wanted it to be decipherable to
all sentient beings, then she would place it on the most cosmic of all
billboards, the Cosmic Microwave Background (CMB) sky.  Here we point out 
that the spherical harmonic coefficients of the observed
CMB anisotropies (or their squared amplitudes at each multipole) depend on
the location of the observer, in both space and time.  The amount of 
observer-independent information available in the CMB is a small fraction 
of the total that any observer can measure.  Hence a lengthy message on the
CMB sky is fundamentally no less observer-specific than a communication hidden
in this morning's tea-leaves.  Nevertheless, the CMB sky {\it does\/}
encode a wealth of information about the structure of the cosmos and possibly
about the nature of physics at the highest energy levels.  The Universe
has left us a message all on its own.
\end{abstract}


\maketitle

\date{today}

\noindent
In a recent posting to the physics arXiv \cite{HsuZee}, Hsu and Zee (HZ) 
discuss the possibility that if a superior Being created a universe, and 
wished to inform the inhabitants of that universe of the existence of the 
creator, then an obvious place to put such a message would be in the 
anisotropies of the Cosmic Microwave Background (CMB, \cite{CMB}).
This would have the apparent
benefit that the encoded information would be readable by all advanced 
civilisations in that universe.  In this short Note we point out that the 
amount of information that the CMB can encode about fundamental physics is 
far less than HZ's estimate, and that almost all of the information in the CMB 
is strongly observer-dependent.  While we agree that precision measurements of 
the CMB anisotropies are a worthwhile endeavour, we stress that there are 
quite Creator-independent reasons for this.

HZ propose that by tweaking the inflaton potential, a superior Being could 
encode a surprisingly large message of $100$ kbits or so in the CMB $C_\ell$ 
spectrum of anisotropies (roughly the size of the raw 
text file for this Note!).  While it is certainly true that the particular 
realisation of the CMB sky that we observe contains a great deal 
of information, almost all of it describes the details 
of the fluctuations within the intersection of our past light cone 
with the last scattering surface, as we explain below.

The amplitude of primordial fluctuations generated by inflation is determined
by the value of the inflaton potential and its slope while the modes are 
leaving the Hubble radius.  However, the connection 
between the $C_\ell$ spectrum and the inflaton potential is not direct 
for at least two important reasons.
Firstly, the usual identification between the inflaton potential and the
fluctuation spectrum $P(k)$ can only be made when the slow roll 
approximation is valid.  One cannot encode an arbitrary pattern in the 
primordial spectrum with a corresponding tweak to the potential.

Another important reason the inflaton potential does not mapped directly 
to the $C_\ell$ spectrum is that the primordial fluctuation spectrum in 
$k$-space undergoes a complicated convolution to form the CMB spectrum in 
$\ell$-space.  The net result is a smoothed version of the $k$-space 
spectrum, which itself does not directly reflect the inflaton potential.  
Therefore HZ's suggestion, based on their estimate of $C_\ell/\Delta 
C_\ell \simeq \sqrt{\ell}$ {\em independent} possible values for each $\ell$ 
(where $\Delta C_\ell$ is the inevitable cosmic variance), greatly
overestimates the amount of information that can be encoded.
No amount of tinkering with the
potential can imprint an arbitrary pattern in the $C_\ell$s -- the 
inherent smoothing results in strong correlations between neighbouring 
$\ell$s.

Furthermore, HZ assume that information can be encoded in the 
CMB spectrum up to $\ell_{\rm max} \sim 10^4$.  In practice, the primary
CMB power is greatly suppressed and smoothed above $\ell \simeq 2000$ due to
the finite thickness of the last scattering surface.  Moreover, due to
foregrounds, in practice one can only observe a fraction of the CMB sky, so
$\Delta C_\ell$ is somewhat higher.
In addition to these problems, we {\em already} have measurements approaching 
cosmic-variance-limited accuracy on the first few hundred $C_\ell$s
\cite{ScottSmoot}, and the measured spectrum is 
consistent with standard $\Lambda$CDM models \cite{except}.
No message exists in this 
range of $\ell$ above the cosmic-variance noise.

It is straightforward to crudely model the effect of the smoothing 
of the $C_\ell$ spectrum to obtain an improved estimate on the total 
information that it can encode.  Suppose that the smoothing has a 
characteristic scale of $\Delta\ell$.  In this case we 
could distinguish very roughly $N$ distinct spectra above the cosmic 
variance, with
\begin{equation}
\ln N \sim \ln \!\! \prod_{i=1}^{\ell_{\rm max}/\Delta\ell} \!\!
\sqrt{i\Delta\ell}
\simeq \frac{\ell_{\rm max}}{2\Delta\ell}\ln(\ell_{\rm max}/e).
\end{equation}
Here the product is restricted to every $\Delta\ell$th value of $\ell$,
and Stirling's approximation was used.  Using the values
$\ell_{\rm max} = 1000$ and $\Delta\ell=10$ \cite{deltal} then gives our
{\em conservative} estimate of $\ln N \sim 300$ bits.

HZ also propose that a superior Being may be advanced enough to 
directly set the spatial phases of the primordial modes, thereby imprinting 
a spatial pattern directly on the $a_{\ell m}$ \cite{alm} coefficients of the
CMB sky.  But this contradicts the 
authors' intent of finding an observer-independent medium for a message.  
This is because the CMB sky is not a static billboard in space, but a 
consequence of the particular set of density variations on the last 
scattering surface, varying (slowly) in time, and different for each observer.

The specific values of the $C_\ell$s in the observed CMB angular power
spectrum, let us call them $a_\ell^2$, arise from a projection of the
3-dimensional pattern of density perturbations on the `last scattering
surface' (a spherically shell, with finite thickness, surrounding every
observer).  In practice there are several contributions to the anisotropies,
but for simplicity let us consider the large angle Sachs-Wolfe effect
\cite{SW} for a flat cosmology:
\begin{equation}
a_\ell^2 \propto  \int (dk/k^2) \delta_k^2 j^2_\ell(k R_{\rm LS}),
\end{equation}
where $\delta_k^2$ is the primordial power spectrum of density perturbations 
and $R_{\rm LS}$ is the distance to last-scattering.
The expectation for the $a_\ell^2$
($C_\ell\equiv\left\langle a_\ell^2\right\rangle$) is a smooth function of
$\ell$, which can be calculated precisely for a given set of initial
conditions (i.e.~$\left\langle \delta_k^2\right\rangle$).  However, 
the actual values of $a_\ell^2$ depend on the specific $\delta_k$s which
are sampled on each observer's last scattering surface.  This means that
different observers (separated by cosmological distances) will observe
different values of the $a_\ell^2$s.  This `cosmic variance' does {\it not\/}
limit the precision with which a set of $a_\ell^2$s can be determined, it
just affects the ability to constrain the underlying expectation values,
i.e.~the $C_\ell$s.

Although one could imagine setting up a set of correlations in the sky
\cite{pi} among $a_{\ell m}$s, it is clearly impossible to arrange for the
same set of correlations to be projected on the last-scattering surface for
{\it all\/} observers, while maintaining anything like statistical isotropy
\cite{cutsky}.

It's even worse than that though, since the $a_{\ell m}$s also change with
the time of observation, through a combination of the 3-dimensional density 
perturbations evolving and the region of space being sampled by the last 
scattering surface moving.  This means that any such communication presented 
on the CMB sky would be a message placed there {\it for us, at the present 
epoch}.

Hence a `message in the sky' is no better than a signal hidden in the
human genome, or indeed a billboard picture of a plate of spaghetti that a
particular set of drivers pass on a specific set of days.

But are there other observer-independent ways of encoding a message?  We have
been unable to come up with any ideas which do not suffer from some of the same
shortcomings as the CMB.
We suggest that it may be impossible to think of any signal which a
supreme Being could imprint on the universe they have created {\it without\/}
it being targetted at particular civilizations or individuals.
Given that faith is an important part of religion, then whether this would be
a proof {\it for\/} or {\it against\/} the existence of an
Intelligent Designer would appear
to be a matter of taste.  Indeed, we would imagine that any Creator would 
operate with greater subtlety than running advertisements on billboards.

HZ's suggestion that a supreme Being might encode the Grand Unified gauge
group in the CMB sky is an interesting one, because in fact there
is real hope that the CMB sky {\it will\/} provide information about
Grand Unification, through observables which depend on the inflaton potential.
The slope of the initial power spectrum, and particularly the amplitude of
the gravity wave contribution (seen through the polarization `B-modes') hold
great promise for probing physics at energy scales $>10^{14}$~GeV.  So there
is indeed a `message in the sky' -- it is the fact that the CMB anisotropies
allow us to determine very precisely the large-scale structure of the
observable Universe \cite{SMC}, and to probe physics at the highest
energies.

\medskip

\noindent\textbf{\large Acknowledgments}
\smallskip

\noindent
We would like to thank the Universe for giving us the CMB \cite{AWord}.


\smallskip



\begin{thebibliography}{9}
\bibitem{alm} The usual definition is
 $T(\theta,\phi)\equiv \sum a_{\ell m} Y_{\ell m}(\theta,\phi)$.  The
 $a_{\ell m}$s are assumed to be Gaussian random numbers with mean zero and
 variance $C_\ell$.
\bibitem{cutsky} And this also ignores the fact that observers living within
galaxies (like us within the Milky Way) will have a significant fraction of
the sky blocked by bright foregrounds.
\bibitem{AWord} This suggests to us the Cosmologist's Anthropic
 Principle, which states: `By existing and doing cosmology
 for a living, we have selected from the multiverse those universes which
 allow us to find out about them.'
\end{thebibliography}
\end{document}